\documentclass[english,prl,groupedaddress,noshowpacs,nofootinbib]{revtex4}
\usepackage[T1]{fontenc}
\usepackage[latin1]{inputenc}
\usepackage{booktabs}
\usepackage{amsmath}
\usepackage{amssymb}

\makeatletter

\providecommand{\tabularnewline}{\\}

\makeatletter

\usepackage{epsfig}

\def\nd{$\nu\phantom{}DE$}
\def\aa{{\cal A}}

\makeatother

\usepackage{babel}
\makeatother

\begin{document}

\title{Neutrino Dark Energy in Grand Unified Theories}

\author{Jitesh R. Bhatt$^{1}$}

\email{jeet@prl.res.in}

\author{Pei-Hong Gu$^{2}$}

\email{pgu@ictp.it}

\author{Utpal Sarkar$^{1}$}

\email{utpal@prl.res.in}

\author{Santosh K. Singh$^{1}$}

\email{santosh@prl.res.in}

\affiliation{$^{1}$Physical Research Laboratory, Ahmedabad 380009, India\\
 $^{2}$The Abdus Salam International Centre for Theoretical Physics,
Strada Costiera 11, 34014 Trieste, Italy}

\begin{abstract}
We studied a left-right symmetric model that can accommodate the neutrino
dark energy (\nd) proposal. Type III seesaw mechanism is implemented
to give masses to the neutrinos. After explaining the model, we study
the consistency of the model by minimizing the scalar potential and
obtaining the conditions for the required vacuum expectation values
of the different scalar fields. This model is then embedded in an
SO(10) grand unified theory and the allowed symmetry breaking scales
are determined by the condition of the gauge coupling unification.
Although $SU(2)_{R}$ breaking is required to be high, its Abelian
subgroup $U(1)_{R}$ is broken in the TeV range, which can then give
the required neutrino masses and predicts new gauge bosons that could
be detected at LHC. The neutrino masses are studied in details in
this model, which shows that at least 3 singlet fermions are required.
\end{abstract}
\maketitle

\section*{Introduction}

During the past couple of decades, astrophysical observations has
improved our knowledge of cosmology tremendously. One of the most
important discovery resulting from these observations is that of the
dark energy\citep{dark}. Nature of the dark energy(DE) is one of
the most puzzling question of physics. The observations suggest that
currently , i.e. around redshift $z\sim1$, the DE is contributing
around $70\%$ of the total energy budget of the universe, while its
contribution was sub-dominant in the past($z\gg1$). Any proposed
model of DE is required to satisfy these observational constraints.
These models require the mass of the scalar to be very light having
scale same as Hubble scale ($\sim10^{-33}eV$). There exist myriad
of such models describing the nature and the dynamics of DE (for recent
reviews see Ref. \citep{review}]. One of the very interesting proposal
for the DE is based on the fact that typical energy scale of DE $\rho_{\Lambda}\sim(3\times10^{-3}\,\textrm{eV})^{4}$
also coincides with the neutrino mass scale $\rho_{\Lambda}\sim m_{\nu}^{4}$.
This has led to several attempts to relate the origin of the dark
energy with the neutrino masses \citep{quint,mavans1,mavans2,mavans3}
and this connection can have many interesting consequences \citep{kap,pas}.
In this scenario a scalar field $\aa$ called the acceleron couples
with the neutrinos and consequently making the neutrino mass $m_{\nu}$
function of $\aa$. Next, it is assumed that the dark energy $\rho_{DE}$
can be written as \[
\rho_{DE}\,=\,\rho_{\nu}\,+\, V(\aa).\]
 Stationary condition on $\rho_{DE}$ then lead to varying the neutrino
mass. These type of models are called mass varying neutrino (MaVaN)
models \citep{mavans1,mavans2,mavans3}. In a typical MaVaN scenario,
the standard model is extended by including singlet right-handed neutrinos
$N_{i},i=1,2,3$, and giving a Majorana mass to the neutrinos which
varies with $\phi_{a}$. At present our understanding of MaVaN models
is far from being complete, several problems regarding nature origin
and nature of the acceleron field, about its stability \citep{mavans2,azk2005}
etc. continue to remain. There has been a significant progress in
solving some of these problems in the subsequent works \citep{models,tripnd},
but much more needs to be done before this idea could be considered
as a realistic one.

Considering the difficulties involved in constructing a reasonable
MaVaNs model, most of the earlier models restricted themselves to
start with the standard model and include a singlet right-handed neutrino,
or else, include a triplet Higgs scalar. Some time back we constructed
a left-right symmetric model with right-handed neutrinos and type-III
seesaw neutrino masses, which could explain the dark energy with MaVaNs
\citep{san}. In this article we work out some of the details of that
model and embed the model in a grand unified theory. The most important
feature of this model is that the model justifies the smallness of
the very low scale, entering in this model. We have analyzed the consistency
of the problem by minimizing the scalar potential and then have found
the conditions for the required minima that explains the required
mass scales in this MaVaNs model. We also study the gauge coupling
unification in the SO(10) GUT, in which this model has been embedded.
The neutrino masses have also been studied and some conditions on
the number of the singlet fermions have been worked out.

\section*{The Model}

One of the problems with the original MaVaNs is that the condition
from naturalness requires the Majorana masses of the right-handed
neutrinos, which varies with the acceleron field, to be in the range
of eV. In such models of type I seesaw, the model becomes void of
any seesaw, since the smallness of the neutrino masses can not be
attributed to any large lepton number violating scale. Another restrictions
of this model is that the model cannot be embedded in any left-right
symmetric extension of the standard model, because the equal treatment
of the left-handed and the right-handed fields would imply that if
neutrino masses vary with the value of the acceleron field, the charged
fermion masses would also vary and that would relate the scale of
dark energy to the top quark mass scale, which is unacceptable. Although
the constraint from the naturalness condition can be softened in the
\nd models with triplet Higgs scalars \citep{tripnd}, this cannot
be embedded in a left-right symmetric model. We consider here a left-right
symmetric model, where the neutrino masses originate from double seesaw
or type III seesaw mechanism and then show how this model can be embedded
in a grand unified theory.

In the left-right symmetric models, the standard model gauge group
is extended to a left-right symmetric gauge group \citep{lr}, $G_{LR}\equiv SU(3)_{c}\times SU(2)_{L}\times SU(2)_{R}\times U(1)_{B-L}$,
so that the electric charge is defined in terms of the generators
of the group as: \begin{equation}
Q=T_{3L}+T_{3R}+{\frac{B-L}{2}}=T_{3L}+Y\,.\label{1}\end{equation}
 The quarks and leptons transform under the left-right symmetric gauge
group as: \begin{eqnarray}
Q_{L}=\begin{pmatrix}u_{L}\\
d_{L}\end{pmatrix}\equiv[3,2,1,{\frac{1}{3}}] & , & Q_{R}=\begin{pmatrix}u_{R}\\
d_{R}\end{pmatrix}\equiv[3,1,2,{\frac{1}{3}}]\,,\nonumber \\
\ell_{L}=\begin{pmatrix}\nu_{L}\\
e_{L}\end{pmatrix}\equiv[1,2,1,{-1}] & , & \ell_{R}=\begin{pmatrix}N_{R}\\
e_{R}\end{pmatrix}\equiv[1,1,2,-1]\,,\nonumber \\
S_{R}\equiv[1,1,1,0]\,. & ~\end{eqnarray}
 The right-handed neutrinos $N_{R}$ is present in all the left-right
symmetric model, which is dictated by the structure of the fermion
representations and the gauge group. However, in models with type
III seesaw mechanism for neutrino masses one introduces an additional
singlet fermion $S_{R}$. As the name left-right symmetric model,
the model Lagrangian is invariant under the left-right parity transformation
given as: \begin{eqnarray*}
SU(2)_{L} & \leftrightarrow & SU(2)_{R}\,,\\
Q_{L} & \leftrightarrow & Q_{R}\,,\\
\ell_{L} & \leftrightarrow & \ell_{R}\,.\end{eqnarray*}
 In this model we have introduced the singlet field $S_{R}$, but
there is no $S_{L}$. But still this model is consistent with left-right
parity operation, since the field $S_{R}$ transform to its $CP$
conjugate state under the left-right parity as: $S_{R}\leftrightarrow{S^{c}}_{L}$.
This also ensures that the Majorana mass term is invariant under the
parity transformation, because this field $S_{R}$ transform under
the transformation $SU(2)_{L}\leftrightarrow SU(2)_{R}$ to itself
$S_{R}\equiv(1,1,1,0)\leftrightarrow(1,1,1,0)$.

The gauge boson (excluding gluons) sector consist of two triplet and
one singlet as :

\[
W_{\mu L}=\left(\begin{array}{c}
W_{L\mu}^{+}\\
W_{L\mu}^{0}\\
W_{L\mu}^{-}\end{array}\right)\equiv(1,3,1,0),~W_{\mu R}=\left(\begin{array}{c}
W_{R\mu}^{+}\\
W_{R\mu}^{0}\\
W_{R\mu}^{-}\end{array}\right)\equiv(1,1,3,0),~B_{\mu(B-L)}\equiv(1,1,1,0)\]
 There exists several choices of the Higgs scalars, and hence, the
choices of symmetry breaking chain. In the present model, the content
of the Higgs sector will be chosen according to the following desired
symmetry breaking pattern\citep{v}: \begin{eqnarray*}
 &  & SU(3)_{c}\times SU(2)_{L}\times SU(2)_{R}\times U(1)_{(B-L)}~\,\,\,\,\,\,\,\,\,[G_{3221D}]\\
 &  & \,\,\,\,\stackrel{M_{R}}{\rightarrow}SU(3)_{c}\times SU(2)_{L}\times U(1)_{R}\times U(1)_{(B-L)}~~~[G_{3211}]\\
 &  & \,\,\,\,\stackrel{m_{r}}{\rightarrow}SU(3)_{c}\times SU(2)_{L}\times U(1)_{Y}~~~~~~~~~~~~~~~~~~~~\,[G_{321}]\\
 &  & \,\,\,\,\stackrel{m_{W}}{\rightarrow}SU(3)_{c}\times U(1)_{Q}~~~~~~~~~~~~~~~~~~~~~~~~~~~~~~~~~~[G_{em}]\,.\end{eqnarray*}
 Breaking of the left-right symmetric group to $G_{3211}$ requires
a right triplet Higgs scalars $\Delta_{R}$ transforming as $\Delta_{R}\equiv(1,1,3,0)$.
The triplet does not change the rank of the gauge group and only breaks
$SU(2)_{R}\to U(1)_{R}$. Since it does not carry any $U(1)_{B-L}$
quantum number, it cannot give any Majorana masses to the neutral
fermions. For the next symmetry breaking stage, $U(1)_{R}\times U(1)_{B-L}\to U(1)_{Y}$,
we introduce an $SU(2)_{R}$ doublet Higgs scalar field $\chi_{R}\equiv(1,1,2,1)$
\citep{doub1,albright}. The $vev$ of $\chi_{R}$ could also break
$[G_{3221D}]\to[G_{321}]$, if the field $\Delta_{R}$ were not present.
The left-right parity would then require the existence of the fields
$\Delta_{L}\equiv(1,3,1,0)$ and $\chi_{L}\equiv(1,2,1,1)$. Finally,
the standard model symmetry breaking is mediated by a bi-doublet field
$\Phi\equiv(1,2,2,0)$, like in any other left-right symmetric model.
This field has the Yukawa interaction with the standard model fermions
and provide Dirac masses to all of them. We shall introduce one more
Higgs bi-doublet scalar $\Psi\equiv(1,2,2,0)$ that is needed for
the purpose of our model. We also introduce another singlet scalar
field $\eta\equiv(1,1,1,0)$, which acquires a tiny $vev$ of the
order of the light neutrino masses and generate the mass scale for
the dark energy naturally.

Now we write down the explicit forms of all the scalar fields in terms
of their components as

\begin{eqnarray*}
\Delta_{L}=\left(\begin{array}{cc}
\Delta_{L}^{0} & \Delta_{L}^{+}\\
\Delta_{L}^{-} & -\Delta_{L}^{0}\end{array}\right) & , & \Delta_{R}=\left(\begin{array}{cc}
\Delta_{R}^{0} & \Delta_{R}^{+}\\
\Delta_{R}^{-} & -\Delta_{R}^{0}\end{array}\right)\,,\\
\\\Phi=\left(\begin{array}{cc}
\phi_{1}^{0} & \phi_{1}^{+}\\
\phi_{2}^{-} & \phi_{2}^{0}\end{array}\right) & , & \Psi=\left(\begin{array}{cc}
\psi_{1}^{0} & \psi_{1}^{+}\\
\psi_{2}^{-} & \psi_{2}^{0}\end{array}\right)\,,\\
\\\chi_{L}=\left(\begin{array}{c}
\chi_{L}^{+}\\
\chi_{L}^{0}\end{array}\right) & , & \chi_{R}=\left(\begin{array}{c}
\chi_{R}^{+}\\
\chi_{R}^{0}\end{array}\right)\,,\end{eqnarray*}
 The most general scalar potential has to be constructed in such a
way that they respect the left-right parity transformation of the
scalar fields listed below: \begin{eqnarray*}
\chi_{L}\leftrightarrow\chi_{R} & , & \Delta_{L}\leftrightarrow\Delta_{R}\\
\Phi\leftrightarrow\Phi^{\dagger} & , & \Psi\leftrightarrow\Psi^{\dagger}\\
. &  & \eta\leftrightarrow\eta\,.\end{eqnarray*}
 Under the left-right gauge group transformation, the Higgs fields
transform as \begin{eqnarray*}
\Delta_{L}\rightarrow U_{L}~\Delta_{L}~U_{L}^{\dagger} & , & \Delta_{R}\rightarrow U_{R}~\Delta_{R}~U_{R}^{\dagger}\\
\Phi\rightarrow U_{L}~\Phi~U_{R}^{\dagger} & , & \Psi\rightarrow U_{L}~\Psi~U_{R}^{\dagger}\\
\chi_{L}\rightarrow U_{L}~\chi_{L} & , & \chi_{R}\rightarrow U_{R}~\chi_{R}\\
 &  & \eta\rightarrow\eta\,.\end{eqnarray*}

In order to write down the scalar potential we also construct the
fields $\tau^{2}\Phi^{\ast}\tau^{2}$ and $\tau^{2}\Psi^{\ast}\tau^{2}$
from $\Phi$ and $\Psi$ which transform in the same ways as $\Phi$
and $\Psi$. For convenience, we represent $\Phi$ as $\phi_{1}$,
$\tau^{2}\Phi\tau^{2}$ as $\phi_{2}$ (and similarly for $\Psi$)
from now on.

\section*{Potential Minimization}

We first write down the most general renormalizable gauge invariant
scalar potential respecting left-right parity and study details of
potential minimization. Besides left-right parity, we impose following
$Z_{4}$ symmetry on only the Higgs potential to avoid few undesired
terms \begin{equation}
\begin{array}{ccccc}
\chi_{L}\rightarrow i\chi_{L}\,,\, & \chi_{R}\rightarrow-i\chi_{R}\,,\\
\Delta_{L}\rightarrow-\Delta_{L}\,, & \Delta_{R}\rightarrow-\Delta_{R}\,,\\
\Phi\rightarrow\Phi\,\,\,, & \Psi\rightarrow-\Psi\,,\\
 & \eta\rightarrow\eta\,.\end{array} \label{Z4}\end{equation}

 We write the the Higgs potential as a sum of of various parts and
write down each part separately as:
 \begin{eqnarray*}
V & = & ~V_{\phi}+V_{\psi}+V_{\Delta}+V_{\eta}+V_{\chi}+V_{\Delta\phi\psi}+V_{\chi\phi\psi}+V_{\eta\chi\Delta\phi\psi}\\
\\V_{\phi} & = & -\sum_{i,j}\frac{\mu_{\phi ij}^{2}}{2}~tr(\phi_{i}^{\dagger}\phi_{j})+\sum_{i,j,k,l}\frac{\lambda_{\phi ijkl}}{4}~tr(\phi_{i}^{\dagger}\phi_{j})~tr(\phi_{k}^{\dagger}\phi_{l})\\
 &  & +\sum_{i,j,k,l}\frac{\Lambda_{\phi ijkl}}{4}~tr(\phi_{i}^{\dagger}\phi_{j}\phi_{k}^{\dagger}\phi_{l})\\
 &  & ~\\
V_{\psi} & = & -\sum_{i,j}\frac{\mu_{\psi ij}^{2}}{2}~tr(\psi_{i}^{\dagger}\psi_{j})+\sum_{i,j,k,l}\frac{\lambda_{\psi ijkl}}{4}~tr(\psi_{i}^{\dagger}\psi_{j})~tr(\psi_{k}^{\dagger}\psi_{l})\\
 &  & +\sum_{i,j,k,l}\frac{\Lambda_{\psi ijkl}}{4}~tr(\psi_{i}^{\dagger}\psi_{j}\psi_{k}^{\dagger}\psi_{l})\\
\\V_{\Delta} & = & -\frac{\mu_{\Delta}^{2}}{2}~[tr(\Delta_{L}\Delta_{L})+tr(\Delta_{R}\Delta_{R})]+\frac{\lambda_{\Delta}}{4}~[tr(\Delta_{L}\Delta_{L})^{2}+tr(\Delta_{R}\Delta_{R})^{2}]\\
 &  & +\frac{\Lambda_{\Delta}}{4}~[tr(\Delta_{L}\Delta_{L}\Delta_{L}\Delta_{L})+tr(\Delta_{R}\Delta_{R}\Delta_{R}\Delta_{R})]\\
 &  & +\frac{g_{\Delta}}{2}~[tr(\Delta_{L}\Delta_{L})~tr(\Delta_{R}\Delta_{R})]\\
\\V_{\eta} & = & \frac{M_{\eta}^{2}}{2}~\eta^{2}+\frac{\lambda_{\eta}}{4}~\eta^{4}\\
\\V_{\chi} & = & -\frac{\mu_{\chi}^{2}}{2}~[\chi_{L}^{\dagger}\chi_{L}+\chi_{R}^{\dagger}\chi_{R}]+\frac{\lambda_{\chi}}{4}~[(\chi_{L}^{\dagger}\chi_{L})^{2}+(\chi_{R}^{\dagger}\chi_{R})^{2}]\\
 &  & +\frac{g_{\chi}}{2}~[\chi_{L}^{\dagger}\chi_{L}~\chi_{R}^{\dagger}\chi_{R}]\\
\\V_{\Delta\phi\psi} & = & \sum_{i,j}\alpha_{\phi ij}~[\Delta_{L}\Delta_{L}+\Delta_{R}\Delta_{R}]~{tr(\phi_{i}^{\dagger}\phi_{j})}\\
 &  & +\sum_{i,j}\alpha_{\psi ij}~[\Delta_{L}\Delta_{L}+\Delta_{R}\Delta_{R}]~{tr(\psi_{i}^{\dagger}\psi_{j})}\\
 &  & +\sum_{i,j}\beta_{\phi ij}~[~tr(\Delta_{L}\Delta_{L}\phi_{i}\phi_{j}^{\dagger})+tr(\Delta_{R}\Delta_{R}\phi_{i}^{\dagger}\phi_{j})]\\
 &  & +\sum_{i,j}\beta{}_{\psi ij}~[~tr(\Delta_{L}\Delta_{L}\psi_{i}\psi_{j}^{\dagger}+tr(\Delta_{R}\Delta_{R}\psi_{i}^{\dagger}\psi_{j})]\\
 &  & +\sum_{i,j}h_{\Delta\phi ij}~tr(\Delta_{L}\phi_{i}\Delta_{R}\phi_{j}^{\dagger})+\sum_{i,j}h_{\Delta\psi ij}tr(\Delta_{L}\psi_{i}\Delta_{R}\psi_{j}^{\dagger})\\
\\V_{\chi\phi\psi} & = & \sum_{i,j}h_{\phi\chi ij}~[\chi_{L}^{\dagger}\chi_{L}+\chi_{R}^{\dagger}\chi_{R}]~{tr(\phi_{i}^{\dagger}\phi_{j})}\\
 &  & +\sum_{i,j}h_{\psi\chi ij}~[\chi_{L}^{\dagger}\chi_{L}+\chi_{R}^{\dagger}\chi_{R}]~{tr(\psi_{i}^{\dagger}\psi_{j})}\\
\\V_{\eta\chi\Delta\phi\psi} & = & \left(h_{\eta\chi}~[\chi_{L}^{\dagger}\chi_{L}+\chi_{R}^{\dagger}\chi_{R}]+h_{\eta\Delta}~[tr(\Delta_{L}\Delta_{L})+tr(\Delta_{R}\Delta_{R})]\right)~\eta^{2}\\
 &  & +\left(\sum_{i,j}h_{\eta\phi ij}~tr(\phi_{i}^{\dagger}\phi_{j})+\sum_{i,j}h_{\eta\psi ij}~tr(\psi_{i}^{\dagger}\psi_{j})\right)~\eta^{2}\\
 &  & +\sum_{i,j}h_{\eta ij}~\eta~[tr(\phi_{i}^{\dagger}\Delta_{L}\psi_{j})+tr(\phi_{i}\Delta_{R}\psi_{j}^{\dagger})+h.c.]\\
 &  & +\sum_{i}h_{\chi i}~\eta~[\chi_{L}^{\dagger}\phi_{i}\chi_{R}+h.c.]\,.
\end{eqnarray*}

We parametrize the true minima of the potential by giving vacuum expectation
values to different scalar fields as follows.

\begin{eqnarray*}
\phi_{1}=\left(\begin{array}{cc}
v & 0\\
0 & v'\end{array}\right),\,\phi_{2}=\left(\begin{array}{cc}
v' & 0\\
0 & v\end{array}\right)\,, &  & \psi_{1}=\left(\begin{array}{cc}
w & 0\\
0 & w'\end{array}\right),\,\,\,\,\psi_{2}=\left(\begin{array}{cc}
w' & 0\\
0 & w\end{array}\right)\,,\\
\chi_{L}=\left(\begin{array}{c}
0\\
v_{L}\end{array}\right),\,\,\,\,\,\chi_{R}=\left(\begin{array}{c}
0\\
v_{R}\end{array}\right)\,,\, &  & \Delta_{L}=\left(\begin{array}{cc}
u_{L} & 0\\
0 & -u_{L}\end{array}\right),\,\,\,\Delta_{R}=\left(\begin{array}{cc}
u_{R} & 0\\
0 & -u_{R}\end{array}\right)\,\\
\eta=u\,.\end{eqnarray*}

Since the phenomenological consistency requires $v\gg v'$and $w\gg w'$,
we ignore potential terms involving $v'$and $w'$ and write down
the general scalar potential in terms of vacuum expectation values
of different scalar fields

\begin{eqnarray*}
V & = & -\frac{\mu_{\phi}^{2}}{2}~v^{2}+\frac{\lambda_{\phi}}{4}~v^{4}-\frac{\mu_{\psi}^{2}}{2}~w^{2}+\frac{\lambda_{\psi}}{4}~w^{4}\\
 &  & -\frac{\mu_{\Delta}^{2}}{2}~(u_{L}^{2}+u_{R}^{2})+\frac{\lambda_{\Delta}}{4}~\left(u_{L}^{4}+u_{R}^{4}\right)\\
 &  & +\frac{M_{\eta}^{2}}{2}~u^{2}+\frac{\lambda_{\eta}}{4}~u^{4}\\
 &  & -\frac{\mu_{\chi}^{2}}{2}~(v_{L}^{2}+v_{R}^{2})+\frac{\lambda_{\chi}}{4}~(v_{L}^{4}+v_{R}^{4})+\frac{g_{\chi}}{2}~(v_{L}^{2}~v_{R}^{2})\\
 &  & +[(\alpha_{\phi}+\beta_{\phi})v^{2}+(\alpha_{\psi}+\beta_{\psi})w^{2}]~(u_{L}^{2}+u_{R}^{2})+(h_{\Delta\phi}v^{2}+h_{\Delta\psi}w^{2})~u_{L}u_{R}\\
 &  & +(h_{\phi\chi}v^{2}+h_{\psi\chi}w^{2})~(v_{L}^{2}+v_{R}^{2})\\
 &  & +[h_{\eta\chi}(v_{L}^{2}+v_{R}^{2})+h_{\eta\Delta}(u_{L}^{2}+u_{R}^{2})+h_{\eta\phi}v^{2}+h_{\eta\psi}w^{2}]~u^{2}\\
 &  & +h_{\eta}~u(u_{L}+u_{R})vw+h_{\chi}~u(v_{L}v_{R})v\,.\end{eqnarray*}
 For convenience, we have replaced $\lambda_{\phi}+\Lambda_{\phi}\rightarrow\lambda_{\phi},~~\lambda_{\psi}+\Lambda_{\psi}\rightarrow\lambda_{\psi},~~\lambda_{\Delta}+\Lambda_{\Delta}\rightarrow\lambda_{\Delta}$.
The minimization of the potential is studied by taking partial derivatives
with respect to $vev$s of all Higgs fields and then separately equating
them to zero. Solving all such equations will provide us the desired
values. One of the minimization conditions $v_{L}\left(\frac{\partial V}{\partial v_{R}}\right)-v_{R}\left(\frac{\partial V}{\partial v_{L}}\right)=0$
leads to the following relation between $v_{L}$ and $v_{R}$: \[
(v_{R}^{2}-v_{L}^{2})~\left[(\lambda_{\chi}-g_{\chi})v_{L}v_{R}-h_{\chi}uv\right]=0\,.\]
 Since $(v_{R}^{2}=v_{L}^{2})$ is not desirable phenomenologically,
we chose \begin{equation}
v_{L}v_{R}=\frac{h_{\chi}uv}{(\lambda_{\chi}-g_{\chi})}\,.\label{vlvr1}\end{equation}
 Using above relation in an another minimization condition $v_{L}\left(\frac{\partial V}{\partial v_{R}}\right)+v_{R}\left(\frac{\partial V}{\partial v_{L}}\right)=0$,
we get \begin{equation}
v_{L}^{2}+v_{R}^{2}=-\frac{\mu_{\chi}^{2}}{\lambda_{\chi}}\,.\label{eq:vlvr2}\end{equation}
 Parametrizing $v_{L}=A~\sin\theta$, $v_{R}=A~\cos\theta$ and putting
them in the two equations \ref{vlvr1} and \ref{eq:vlvr2} , we find
$A=-\mu_{\chi}^{2}/\lambda_{\chi}$ $\sin2\theta=2\theta=\frac{2h_{\chi}uv}{(\lambda_{\chi}-g_{\chi})}$
since $\mu_{\chi}$ is a large number compared to the numerator. So
we get \begin{eqnarray*}
v_{R} & = & A=\sqrt[2]{-\mu_{\chi}^{2}/\lambda_{\chi}}\,,\\
v_{L} & = & A\theta=\frac{\lambda_{\chi}h_{\chi}}{(g_{\chi}-\lambda_{\chi})}~\frac{uvv_{R}}{\mu_{\chi}^{2}}\,.\end{eqnarray*}
 We have chosen the parametrization of $v_{L}$ and $v_{R}$ in such
a way that $v_{R}$ gets value equal to breaking scale of $G_{3211}$
and $v_{L}$ gets a very small value. We could have done other way
around but that is not what is phenomenologically allowed. Proceeding
with the same kind of analysis for $u_{L}$ and $u_{R}$, i.e., using
two minimization conditions $u_{L}\left(\frac{\partial V}{\partial u_{R}}\right)-u_{R}\left(\frac{\partial V}{\partial u_{L}}\right)=0$
and $u_{L}\left(\frac{\partial V}{\partial u_{R}}\right)+u_{R}\left(\frac{\partial V}{\partial u_{L}}\right)=0$,
we get \begin{eqnarray*}
u_{R} & = & \sqrt[2]{-\mu_{\Delta}^{2}/\lambda_{\Delta}}\,,\\
u_{L} & = & \frac{\lambda_{\Delta}h_{\Delta}}{(g_{\Delta}-\lambda_{\Delta})}~\frac{(h_{\Delta\phi}v^{2}+h_{\Delta\psi}w^{2})u_{R}}{\mu_{\Delta}^{2}}\,.\end{eqnarray*}

Now using equation \ref{vlvr1}, the $\eta$ field can be shown to
get $vev$ only by term $h_{\eta}u(u_{L}+u_{R})$ as only this term
is linear in $u$. The term $h_{\chi}u(v_{L}v_{R})v$ does not remain
linear in $u$ after we substitute the value of $v_{L}v_{R}$ from
equation \ref{vlvr1}. Since the mass term for $\eta$ field is large
and positive, we expect very small $vev$. So we can ignore some of
the terms in the potential while solving for $u$ and can easily obtain
\[
u=\frac{h_{\eta}vw(u_{L}+u_{R})}{M_{\eta}^{2}-(h_{\eta\Delta}\mu_{\Delta}^{2}/\lambda_{\Delta})-(h_{\eta\chi}\mu_{\chi}^{2}/\lambda_{\chi})}\,.\]
 After analyzing the complete scalar potential, we find a consistent
solution with ordering \begin{eqnarray}
u_{R}\gg v_{R}>v>w\gg u\gg v_{L}\,.\phantom{pushhard}\end{eqnarray}
 At this stage we can assume the different mass scales to explain
the model. However, when we embed this model in an $SO(10)$ grand
unified theory, the gauge coupling unification will impose strong
constraints on the different symmetry breaking scales. The left-right
parity and the $SU(2)_{R}$ breaking scale will come out to be above
$10^{11}$~GeV. So, we shall assume $u_{R}\sim10^{11}$~GeV. We
also assume $m_{\eta}\sim m_{\Delta}\sim u_{R}$. However, it will
be possible to keep the $G_{3211}$ symmetry breaking scale to be
very low, and hence, we shall assume $m_{\chi}\sim v_{R}\sim$~TeV.
We find the remaining mass scales to be $v\sim m_{w}\sim100$~GeV,
$u\sim u_{L}\sim$~eV and $v_{L}\sim10^{-2}$~eV.

\section*{Embedding The Model In $SO(10)$GUT}

The idea of Grand Unified Theories (GUTs) has emerged as a very attractive
idea to go beyond Standard Model (SM) for last three decades. It unifies
the three different looking gauge coupling constants of the SM, and
in addition, reduces the number of particle irreducible multiplets
into lesser number of multiplets. The ad-hoc looking hypercharge assignment
in SM gets a predictive framework in GUTs, i.e, the charge quantization
remains no more a surprise in GUTs. The smallest GUT $SU(5)$, in
its non-supersymmetric version, does not unify the three gauge coupling
constants. Out of the higher rank gauge groups containing SM gauge
group as a subgroup, the rank four semi-simple group $SO(10)$ has
emerged as a very attractive candidate for GUT. It can accommodate
the entire SM fermion content in its single 16-dimensional complex
irreducible spinor representation including the right handed neutrino
with three copies for the three families. Its all irreducible representations
are anomaly free providing a natural predictive framework to understand
the fermion masses and mixing. Also the seesaw structure gets a natural
embedding in $SO(10)$. The left-right symmetry group can also be
embedded in $SO(10)$ GUT.

We shall study here the embedding of the present model with all its
Higgs content in $SO(10)$ GUT. We consider the following breaking
pattern of $SO(10)$ gauge group to first Pati-Salam gauge group $SU(4)\times SU(2)_{L}\times SU(2)_{R}$,
next to the left-right gauge group and then to the SM gauge group

\begin{eqnarray*}
SO(10) & \stackrel{M_{U}}{\rightarrow} & SU(4)\times SU(2)_{L}\times SU(2)_{R}\,\,\,\,\,\,\,\,\,\,\,\,\,\,\,\,\,[G_{422D}]\\
 & \stackrel{M_{C}}{\rightarrow} & SU(3)_{c}\times SU(2)_{L}\times SU(2)_{R}\times U(1)_{(B-L)}~[G_{3221D}]\\
 & \stackrel{M_{R}}{\rightarrow} & SU(3)_{c}\times SU(2)_{L}\times U(1)_{R}\times U(1)_{(B-L)}~~~[G_{3211}]\\
 & \stackrel{m_{r}}{\rightarrow} & SU(3)_{c}\times SU(2)_{L}\times U(1)_{Y}~~~~~~~~~~~~~~~~~~~~\,[G_{321}]\\
 & \stackrel{m_{W}}{\rightarrow} & SU(3)_{c}\times U(1)_{Q}~~~~~~~~~~~~~~~~~~~~~~~~~~~~~~~~~~[G_{em}]\,.\end{eqnarray*}
 The Higgs multiplets which can provide the masses for all the SM
fermions are limited as $16\times16=10_{s}+120_{a}+\overline{126}_{s}$.
The 10 dimensional Higgs field $H_{\Phi}$ decomposes under left-right
gauge group as \[
H_{\Phi}\left(10\right)=\Phi(1,2,2;0)\oplus(3,1,1;-\frac{1}{3})\oplus(\overline{3},1,1;\frac{1}{3})\,.\]
 One can easily identify the bi-doublet $\Phi(1,2,2;0)$ appearing
in the left-right model contained in $H_{\Phi}(10)$. To include another
bi-doublet $\Psi(1,2,2;0)$ present in the model, a second Higgs field
$H_{\Psi}(10)$.

Although the fermion and gauge sector of the $SO(10)$ GUT model are
quite simple, the Higgs sector is quite complicated since it is not
only required for generating fermion Masses, but an appropriate Higgs
content is also needed for systematic and consistent breaking of the
$SO(10)$ gauge group down to the SM gauge group in one or more steps.
To break $SO(10)$ gauge group to the Pati-Salam gauge group, one
requires Higgs field either $S(54)$ or $\Upsilon(210)$, which decompose
under Pati-Salam group as \begin{eqnarray*}
S(54) & = & (1,1,1)\oplus(1,3,3)\oplus(20,1,1)\oplus(6,2,2)\,,\\
\Upsilon(210) & = & (1,1,1)\oplus(15,1,1)\oplus(6,2,2)\oplus(15,3,1)\\
 &  & \oplus(15,1,3)\oplus(10,2,2)\oplus(\overline{10},2,2)\,.\end{eqnarray*}
 Giving $vev$ to either of the two fields in the singlet direction
will serve the purpose of the desired breaking. The $(15,1,1)$ of
$\Upsilon$ also has a singlet under the left-right gauge group which
can acquire $vev$ to break the Pati-Salam group to the left-right
group. The $(15,3,1)$ and $(15,1,3)$ Higgs multiplets of $\Upsilon$
also contain the fields $\Delta_{L}(1,3,1,0)$ and $\Delta_{R}(1,1,3,0)$
present the left-right model. Hoever, the $\Upsilon$ singlet under
Pati-salam gauge group is odd under D-Parity. If we give $vev$ to
$\Upsilon$ singlet, the left-right symmetry will be broken at unification
scale itself. Since our model is left-right symmetric, we must avoid
D-parity breaking until left-right group is broken.

However, the singlet in $S(54)$ field under Pati-Salam gauge group
does respect and so can be used to break the GUT group to the Pati-Salam
gauge group. But, the breaking with $S(54)$ does not serve the purpose
of further breaking to the left-right group. So for the next step
breaking, a Higgs Field $A(45)$ is needed along with $S(54)$ which
has the decomposition under the left-right group as

\begin{eqnarray*}
A(45) & = & (1,1,1;0)\oplus\Delta_{L}(1,3,1;0)\oplus\Delta_{R}(1,1,3;0)\\
 &  & \oplus(3,1,1;\frac{4}{3})\oplus(\overline{3},1,1;-\frac{4}{3})\oplus(8,1,1;0)\\
 &  & \oplus(3,2,2;\frac{2}{3})\oplus(\overline{3},2,2;-\frac{2}{3})\,.\end{eqnarray*}
 The first row of the above decomposition is of our interest as it
contains the fields $\Delta_{L}(1,3,1,0)$ and $\Delta_{R}(1,1,3,0)$
of our model along with the left-right group singlet. This singlet
is even under D-parity and so the left-right symmetry is unbroken
until $\Delta_{R}$ acquires $vev$ along the singlet direction to
the SM gauge group. We will be following this approach in the remaining
part of this section.

Now the fields $\chi_{L\,\,}(1,2,2,1)$ and $\chi_{R}(1,1,2,1)$ are
still left to be embedded in some tensors of $SO(10)$. The desired
quantum numbers indicate that they can be embedded in the spinorial
Higgs representation $\left(C(16)\oplus\overline{C(16)}\right)$ .
Decomposition of the $16\oplus\overline{16}$ spinor representation
under left-right group are given as

\begin{eqnarray*}
C(16) & = & \chi_{L}^{\ast}(1,2,1,-1)\oplus\chi_{R}(1,1,2,1)\\
 &  & \oplus(3,2,1,\frac{1}{3})\oplus(\overline{3},1,2,-\frac{1}{3})\,,\\
\overline{C(}\overline{16)} & = & \chi_{L}(1,2,1,1)\oplus\chi_{R}^{\ast}(1,1,2,-1)\\
 &  & \oplus(3,1,2,\frac{1}{3})\oplus(\overline{3},2,1,-\frac{1}{3})\,.\end{eqnarray*}

Having embedded all the Higgs fields of our model into $SO(10)$ tensor
fields, we now write vacuum expectation values along the three singlet
direction under the SM group of the fields $A(45)$ and $S(54)$ as

\begin{eqnarray*}
\left\langle A\right\rangle  & = & M_{C}\hat{A}_{C}+M_{R}\hat{A}_{R}\,,\\
\\\left\langle S\right\rangle  & = & M_{U}\hat{S}\,,\end{eqnarray*}
 where $\hat{A}_{C}$, $\hat{A}_{R}$ and $\hat{S}$ are the singlet
directions under the SM gauge group given as

\begin{eqnarray*}
\hat{A}_{C} & = & \left(\hat{A}_{56}+\hat{A}_{78}+\hat{A}_{910}\right)\\
\hat{A}_{R} & = & \left(\hat{A}_{12}+\hat{A}_{34}\right)\\
\hat{S} & = & 3\times\sum_{a=1}^{4}\hat{S}_{aa}-2\times\sum_{a=5}^{10}\hat{S}_{aa}\,.\end{eqnarray*}
 The indices (1, 2, 3, 4 ) belong to $SO(4)$ and (5, 6, 7, 8, 9,
10) belong to $SO(6)$ subgroup of the group $SO(10)$. We have not
taken care of the normalization factors while writing the directions
of the singlets as they are not much relevant for the present discussion.
However, we can assume that the normalization factors are absorbed
in the corresponding $vev$ values and can proceed without worrying
about them for an approximate analysis. 

Let us denote $H_{\Phi}=h$, $H_{\Psi}=H$ for simplicity in notations.
Now we write the most general $SO(10)$ invariant Higgs potential:

\begin{eqnarray*}
V & = & \mu_{A}^{2}\, A_{ab}A_{ba}+\mu_{S}^{2}\, S_{ab}S_{ba}+\mu_{h}^{2}\, h_{a}h_{a}+\mu_{H}^{2}\, H_{a}H_{a}+\mu_{C}^{2}\,\left(\bar{C}C\right)+\mu_{\eta}^{2}\,\eta^{2}+\lambda_{\eta}\,\eta^{4}\\
 & + & \lambda_{A}\, A^{2}A^{2}+\lambda'_{A}A^{4}+\lambda_{S}\, S^{4}+\lambda_{h}\, h^{4}+\lambda_{H}\, H^{4}+\lambda_{c}\,\left(\bar{C}C\right)^{2}+\lambda_{c}^{'}\,\left(C^{4}+\bar{C}^{4}\right)\\
 & + & g_{AS}A^{2}S^{2}+g'_{AS}A_{ab}A_{bc}S_{cd}S_{da}+g''_{AS}A_{ab}S_{bc}A_{cd}S_{da}\\
 & + & h_{a}\left(g_{hA}A_{ab}A_{bc}+g_{hS}S_{ab}S_{bc}\right)h_{c}+\left(g'_{hA}A^{2}+g'_{hS}S^{2}\right)h^{2}\\
 & + & H_{a}\left(g_{HA}A_{ab}A_{bc}+g_{HS}S_{ab}S_{bc}\right)H_{c}+\left(g'_{HA}A^{2}+g'_{HS}S^{2}\right)H^{2}\\
 & + & \left(g_{hC}h^{2}+g_{HC}H^{2}+g_{AC}A^{2}+g_{SC}S^{2}\right)\,\bar{C}C+g_{\eta HC}\eta\, h\,\left(CC+\bar{C}\bar{C}\right)\,.\end{eqnarray*}
The $Z_{4}$ symmetry (expression \ref{Z4}) used while writing the Higgs potential invariant
under left-right gauge group has also been imposed here on the corresponding
$SO(10)$ Higgs multiplets. Moreover, we have prevented some of the
terms by applying the discrete symmetry $S\rightarrow-S$. The realization
of the first three symmetry breaking steps is possible by taking the
following structure of the $vev$ assignments to the fields $A(45)$
and $S(54)$:

\begin{eqnarray*}
\left\langle A\right\rangle  & = & i\tau_{2}\otimes diag\left(M_{R},\, M_{R},\, M_{C},\, M_{C},\, M_{C}\right)\\
\langle S\rangle & = & I\otimes diag\left(-\frac{3}{2}M_{U},\,-\frac{3}{2}M_{U},\, M_{U},\, M_{U},\, M_{U}\right)\,.\end{eqnarray*}
For the matter of convenience we have just replaced the $vev$s with
the corresponding breaking scales. The potential , in terms of the
$vev$ values of $A$ and $S$, will be approximately given as

\begin{eqnarray*}
V & = & \mu_{A}^{2}\left(6M_{C}^{2}+4M_{R}^{2}\right)+\mu_{S}^{2}\,15M_{U}^{2}+\left(\mu_{C}^{2}+g_{AC}6M_{C}^{2}+g_{SC}\,15M_{U}^{2}\right)\bar{C}C\\
 & + & \left(\mu_{h}^{2}+g_{hS}9M_{U}^{2}\right)h_{a}h_{a}\left(a=1-4\right)+\left(\mu_{h}^{2}+g_{hA}6M_{C}^{2}+g_{hS}6M_{U}^{2}\right)h_{a}h_{a}\left(a=5-10\right)\\
 & + & \left(\mu_{H}^{2}+g_{HS}9M_{U}^{2}\right)H_{a}H_{a}\left(a=1-4\right)+\left(\mu_{H}^{2}+g_{HA}6M_{C}^{2}+g_{HS}6M_{U}^{2}\right)H_{a}H_{a}\left(a=5-10\right)\\
 & + & \lambda_{A}\left(6M_{C}^{2}+4M_{R}^{2}\right)^{2}+\lambda'_{A}\left(6M_{C}^{4}+4M_{R}^{4}\right)+\lambda_{S}M_{U}^{4}+g_{AS}M_{U}^{2}\left(6M_{C}^{2}+9M_{R}^{2}\right)\\
 & + & \lambda_{h}h^{4}+\lambda_{H}H^{4}+g_{\eta HC}\eta\, h\,\left(CC+\bar{C}\bar{C}\right)+\lambda_{c}\,\left(\bar{C}C\right)^{2}+\lambda'_{c}\,\left(C^{4}+\bar{C}^{4}\right)+\lambda_{\eta}\,\eta^{4}\end{eqnarray*}
We have assumed $M_{R}\ll M_{U}\sim M_{C}$ while writing the final
form of the potential. In order to give desired masses (of the order
of $M_{W}$) to the two left-right bi-doublets , $\mu_{h}$ and $\mu_{H}$
will have to be fine-tuned at the order of scale of $M_{U}$. The
fine-tuning can produce very large masses to the triplets of $h$(
or $H$) provided the condition $\left(g_{hA}6M_{C}^{2}-h_{hS}3M_{U}^{2}\right)\sim\left(+M_{U}^{2}\right)$
is satisfied. Another fine-tuning is required in the mass parameter
$\mu_{C}^{2}$ to provide the desired $TeV$ scale masses to the Higgs
fields $C(16)\oplus\overline{C(16)}$. Before ending this section,
we would like to notice an important point. If we take the $g_{SA}$
coupling to be very small, we can argue that the appearance of the
similar combination $\left(6M_{C}^{2}+4M_{R}^{2}\right)$ everywhere
in the potential allows $M_{C}$ and $M_{R}$ to take quite different
values without disturbing other part of the potential. So the scale
of $M_{C}$ and $M_{R}$ can be chosen to be different by orders of
magnitude to get the desirable breaking.

\section*{Gauge Coupling Evolution}

In the present section, we will be studying the set of two-loop renormalization
group (RG) equations for the evolution of the coupling constants and
will be verifying the consistency of the chosen $vev$ for different
Higgs fields in the context of $SO(10)$ GUT. For simplicity, we assume
that the scale $M_{U}$ and $M_{C}$ are very close and we ignore
the  evolution of the coupling constants between the two scales. This
is quite preferable as we will see later that the unification scale
is very tightly constrained by the current proton decay bound \citep{:2009gd}
and any substantial difference between the two breaking scales would
make it even worse. We start with the following equation for the two-loop
evaluation of the coupling constant $\alpha_{i}$

\begin{equation}
\frac{d\alpha_{i}^{-1}\left(t\right)}{dt}=-\frac{a_{i}}{2\pi}-\frac{b_{ij}}{8\pi^{2}}\left(\frac{1}{\alpha_{j}^{-1}}\right)\label{eq:alpha-evolution}\end{equation}
 where $t=\ln\left(M_{\mu}\right)$ and $M_{\mu}$ is the desired
energy scale where the couplings constants, $\alpha_{i}$'s, are be
determined. The $a_{i}$'s and $b_{ij}$'s are the one-loop and
two-loop beta functions governing the evolution of $\alpha_{i}$'s
and include the contributions from gauge bosons, fermions and scalars
in the model.

The fermion contribution to the beta function is taken right from
the starting, the electroweak scale (100GeV). The contributions of
the gauge bosons to beta functions are straightforward to compute
as one can easily determine the expected mass scales of the heavy
gauge bosons corresponding to any given gauge group. However, the
contribution coming from the Higgs content is not so clear because
the heavy Higgs modes can have various possible mass spectrums. We
will use the extended survival hypothesis to fix this uncertainty.
The extended survival hypothesis is based on the assumption that only
minimal number of fine-tunings of the parameters in the Higgs potential
are imposed to ensure the hierarchy in various gauge boson masses.
According to the extended survival hypothesis, only those scalar multiplets
are present at any given intermediate breaking scale $M_{I}$ of a
intermediate gauge group $G_{I}$ which are either required for breaking
the gauge group $G_{I}$ or needed to further break any other intermediate
gauge group below scale $M_{I}.$

A list of Higgs multiplets surviving at the breaking scale of a intermediate
group $G_{I}$, using the extended survival hypothesis, are given
in table. A list of both one-loop and two-loop beta coefficients,
which include all the contributions, that govern the evolution above
the breaking scale of $G_{I}$ to the next intermediate scale are
also listed.

\begin{table}
\begin{tabular}{cccc}
\toprule
Group $G_{I}$ & Higgs content & a & b\tabularnewline
\midrule
\midrule
\addlinespace[3mm]
$G_{321}$ & $\begin{array}{l}
\left(1,\,2,\,\frac{1}{2}\right)_{10}\oplus\left(1,\,2,\,-\frac{1}{2}\right)_{10}\\
\left(1,\,2,\,\frac{1}{2}\right)_{10'}\oplus\left(1,\,2,\,-\frac{1}{2}\right)_{10'}\end{array}$ & $\left(\begin{array}{l}
-7\\
\\-3\\
\\\,\,\frac{21}{5}\\
\\\end{array}\right)$ & $\left(\begin{array}{lll}
-26 & \frac{9}{2} & \frac{11}{10}\\
\\\,12 & 8 & \frac{6}{5}\\
\\\frac{44}{5} & \frac{18}{5} & \frac{104}{25}\end{array}\right)$\tabularnewline\addlinespace[5mm]
\addlinespace[3mm]
$G_{3211}$ & $\begin{array}{l}
\left(1,\,2,\,\frac{1}{2}\,0\right)_{10}\oplus\left(1,\,2,\,-\frac{1}{2}\,0\right)_{10}\\
\left(1,\,2,\,,\frac{1}{2}\,0\right)_{10'}\oplus\left(1,\,2,\,-\frac{1}{2},\,0\right)_{10'}\\
\left(1,\,1,\,-\frac{1}{2},\,\frac{1}{2}\right)_{16}+\left(1,\,1,\,\frac{1}{2},\,-\frac{1}{2}\right)_{\overline{16}}\end{array}$ & $\left(\begin{array}{l}
-7\\
\\-3\\
\\\,\,\frac{53}{12}\\
\\\,\,\frac{33}{8}\end{array}\right)$ & $\left(\begin{array}{llll}
-26 & \frac{9}{2} & \frac{3}{2} & \frac{1}{2}\\
\\\,12 & 8 & 1 & \frac{3}{2}\\
\\\,12 & 3 & \frac{17}{4} & \frac{15}{8}\\
\\\,\,4 & \frac{9}{2} & \frac{15}{8} & \frac{65}{16}\end{array}\right)$\tabularnewline\addlinespace[5mm]
\addlinespace[3mm]
$G_{3221D}$ & $\begin{array}{l}
\left(1,\,2,\,2,\,0\right)_{10}\\
\left(1,\,2,\,2,\,0\right)_{10'}\\
\left(1,\,2,\,1,\,-\frac{1}{2}\right)_{16}\oplus\left(1,\,2,\,1,\,\frac{1}{2}\right)_{\overline{16}}\\
\,\left(1,\,1,\,2,\,\frac{1}{2}\right)_{16}\oplus\left(1,\,1,\,2,\,-\frac{1}{2}\right)_{\overline{16}}\\
\left(1,\,1,\,3,\,0\right)_{45}\\
\left(1,\,3,\,1,\,0\right)_{45}\end{array}$ & $\left(\begin{array}{l}
-7\\
\\-\frac{5}{2}\\
\\-\frac{5}{2}\\
\\\,\,\frac{9}{2}\end{array}\right)$ & $\left(\begin{array}{llll}
-26 & \frac{9}{2} & \frac{9}{2} & \frac{1}{2}\\
\\\,12 & \frac{39}{2} & 3 & \frac{9}{4}\\
\\\,12 & 3 & \frac{39}{2} & \frac{9}{4}\\
\\\,\,4 & \frac{27}{4} & \frac{27}{4} & \frac{23}{4}\\
\\\end{array}\right)$\tabularnewline\addlinespace[5mm]
\bottomrule
\end{tabular}

\caption{Higgs multiplets at different intermediate breaking scales along with
the both one-loop and two-loop beta coefficientss, including all the
contributions from fermions, gauge bosons and Higgs bosons, which
govern the evolution of coupling constants above breaking scale of
$G_{I}$ to the next breakingscale.}

\end{table}

Since our model contains intermediate steps, we require appropriate
matching conditions at the corresponding breaking scales. For the
tow-loop RG running of the coupling constants, the matching conditions
have been derived in \citep{Weinberg:1980wa,Hall:1980kf}. Suppose
a gauge group $G$ is spontaneously broken into a sub-group $\prod_{i}G_{i}$
with several individual factors $G_{i}$, then the following matching
condition need to be satisfied for the two-loop analysis

\begin{equation}
\alpha_{G}^{-1}\left(M_{I}\right)-\frac{C\left(G\right)}{12\pi}=\alpha_{G_{i}}^{-1}\left(M_{I}\right)-\frac{C\left(G_{i}\right)}{12\pi}\,,\label{eq:matching-condition}\end{equation}
where $C(G/G_{i})$ is the quadratic Casimir invariant for the group
$G/G_{i}$. We choose initial starting values of the above three coupling
constants ( central values) at scale $M_{W}$ to be $\alpha_{1Y\text{ }}^{-1}(M_{W})=59.38$,
$\alpha_{2L}^{-1}(M_{W})=29.93$, and $\alpha_{3c}^{-1}(M_{W})=8.47$.
Now let us write the

The boundary conditions at various breaking scales, using the expression
\ref{eq:matching-condition}, can be written as

1. At scale $m_{r}$: \begin{eqnarray*}
\alpha_{1Y\text{ }}^{-1}(m_{r}) & = & \frac{3}{5}\alpha_{1R\text{ }}^{-1}(m_{r})+\frac{2}{5}\alpha_{1(B-L)}^{-1}(m_{r})\,.\end{eqnarray*}

2. At scale $M_{R}$:

\begin{eqnarray*}
\alpha_{1R\text{ }}^{-1}(M_{R}) & = & \alpha_{2R\text{ }}^{-1}(M_{R})-\frac{2}{12\pi}\,,\\
\alpha_{2R\text{ }}^{-1}(M_{R}) & = & \alpha_{2L\text{ }}^{-1}(M_{R})\,.\end{eqnarray*}

3. At the unification scale $M_{U}$\begin{eqnarray*}
\alpha_{2L\text{ }}^{-1}(M_{U})-\frac{2}{12\pi} & = & \alpha_{2R\text{ }}^{-1}(M_{U})-\frac{2}{12\pi}\\
 & = & \alpha_{U}^{-1}\left(M_{U}\right)-\frac{8}{12\pi}\,,\\
\alpha_{3c}^{-1}\left(M_{U}\right)-\frac{3}{12\pi} & = & \alpha_{U}^{-1}\left(M_{U}\right)-\frac{8}{12\pi}\,,\\
\alpha_{B-L}^{-1}\left(M_{U}\right) & = & \alpha_{U}^{-1}\left(M_{U}\right)-\frac{8}{12\pi}\,.\end{eqnarray*}
The matching conditions at the unification scale have been written
by assuming the Pati-Salam scale to be almost close to the unification
scale.

\begin{figure}
\begin{centering}
\epsfig{file=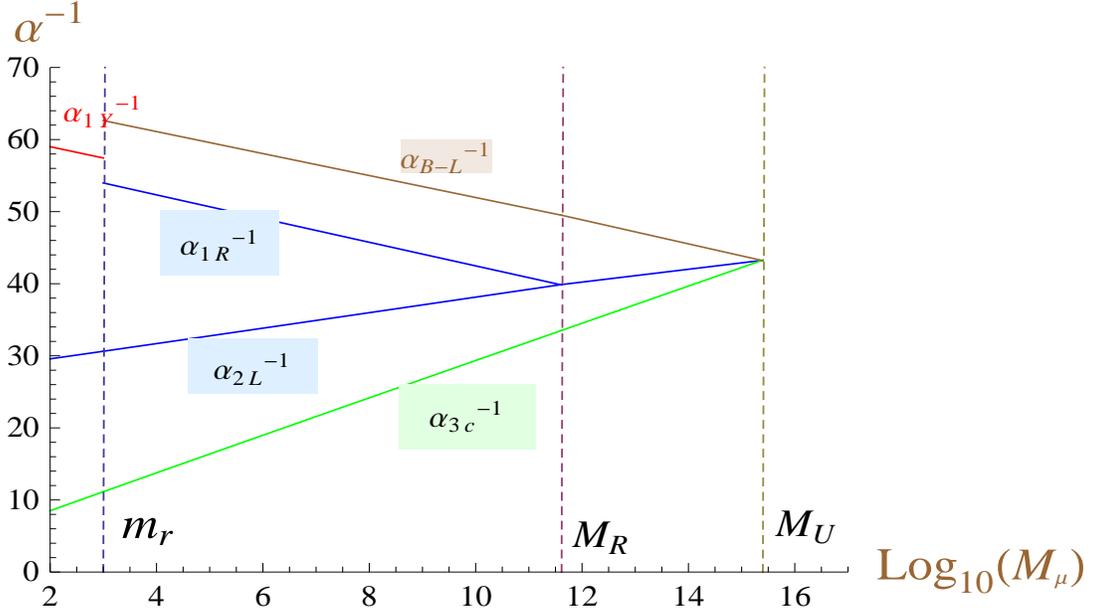, width=0.8\textwidth,height=0.35\textheight}
\par\end{centering}

\caption{\label{fig:Evolution-of-coupling}Evolution of coupling constants}

\label{c-evolution}
\end{figure}

Using the above boundary conditions we have numerically solved the
equation \ref{eq:alpha-evolution} for the two-loop RG evolution for
all the coupling constants. We have taken the breaking scale of the
gauge group $G_{3211}$ to be around $1$TeV. The unification scale
comes out to be $M_{U}=10^{15.4}$GeV and the corresponding coupling
constant is estimated as $\alpha_{U}^{-1}\left(M_{U}\right)=43.4$.
Also the breaking scale of left-right symmetric gauge group, i.e.,
$G_{3221D}$ turns out to be $M_{R}=10^{11.6}$GeV. The running of
the various coupling constants with energy scale are shown in figure
\ref{fig:Evolution-of-coupling}.

However, we find that the scale of the unification along with the
$\alpha_{U}-1$ are not satisfying the most recent bounds on proton
decay, although very close to the limit. The current experimental
lower bound of the partial life time for $p\rightarrow e^{+}\pi^{0}$
is $\tau_{p}>8.2\times10^{33}$ years and for $p\rightarrow\mu^{+}\pi^{0}$
is $\tau_{p}>6.6\times10^{33}$ years \citep{:2009gd}. The theoretical
decay rate of the proton can be estimated as:
\[
\Gamma_{p}\simeq\alpha_{GUT}^{2}\frac{m_{p}^{5}}{M_{X,Y}^{4}}\,.\]

This can be used to estimate the lower limit of the Heavy gauge boson
masses. If the mass scale of super heavy gauge bosons are given as
$M_{X}\simeq10^{n}$GeV, the above proton decay bound is equivalent
to

\begin{equation}
\kappa=\left(\frac{\alpha_{GUT}}{45}\right)\times10^{2(n-15)}\gtrsim11.8\,.\label{eq:proton-bound}\end{equation}
What we obtain for the value of $\kappa$ in our analysis is $\kappa=6.07$.
This is below the lower limit allowed by the proton decay bound as
specified in the right-hand side of the expression \ref{eq:proton-bound}.
However, the value of $\kappa$ is very close to the allowed lower
limit and so we will try to explore the viability of our model by
allowing threshold uncertainty in the Higgs spectrum at various intermediate
breaking scales. It is important to remark at this point that we could
get the reported value of $\kappa$ to be close to the limit only
when we optimized certain degrees of freedom in the Higgs sector.
For instance, the Higgs-bidoublet $\Phi$ has been asumed to arise
from a real 10-dimensional $SO(10)$ Higgs $H_{\Phi}$. So $\Phi$
would not be equivalent to two SM Higgs doublets at the electroweak
scale but will be equivalent to only one such doublet. Similar asuumption
has been also taken for $\Psi$. However, we would like to emphasize
that the results and discussion of the potential minimization will
remain almost same.

\begin{table}
\begin{tabular}{ccc}
\hline
$\begin{array}{l}
\mathrm{SO(10)\, Higgs}\\
\mathrm{Representation}\end{array}$ & $\begin{array}{l}
\mathrm{Higgs\, multiplets\, contributing\,}\\
\mathrm{to\, threshold\, uncertainty}\\
\left(\mathrm{Decomposed\, under}\, G_{3211}\right)\end{array}$ & $\left\{ a_{3c},\, a_{2L},\, a_{1R},\, a_{1(B-L)}\right\} $\tabularnewline
\hline
\hline
\noalign{\vskip3mm}
$16$ & $\begin{array}{l}
\left(1,\,1,\,\frac{1}{2},\,\frac{1}{2}\right)_{16}\oplus\left(1,\,1,\,-\frac{1}{2},\,-\frac{1}{2}\right)_{\overline{16}}\\
\left(1,\,2,\,0,\,-\frac{1}{2}\right)_{16}\oplus\left(1,\,2,\,0,\,\frac{1}{2}\right)_{\overline{16}}\end{array}$ & $\left\{ 0,\,1,\,\frac{1}{2},\,\frac{9}{4}\right\} $\tabularnewline[5mm]
\noalign{\vskip3mm}
$45$ & $\begin{array}{l}
\left(1,\,3,\,1,\,0\right)_{45}\end{array}$ & $\left\{ 0,\,2,\,0,\,0\right\} $\tabularnewline[5mm]
\hline
\end{tabular}

\caption{\label{tab:Threshold_left-right}Threshold contribution at left-right
breaking scale}

\end{table}

The threshold uncertainty in the Higgs spectrum arises form the fact
that the Higgs bosons becoming heavy at a given breaking scale may
not get exactly same masses equal to the energy corresponding to the
breaking scale. However, the Higgs mass spectrum is expected to be
scattered around the energy of the breaking scale within an small
width. For our analysis, we follow a similar approach discussed in
\citep{Mohapatra:1992dx}. We assume that the masses of the Higgs
bosons are scattered around the breaking scale within the factor of
$\frac{1}{30}$ to $30$. So if the mass of a Higgs multiplet at the
given breaking scale $M_{I}$ is $M_{H}$, then we expect \[
\frac{1}{30}\lesssim\frac{M_{H}}{M_{I}}\lesssim30\,.\]

To include the threshold uncertainty at a given breaking scale, we
need to slightly modify our matching conditions at that scale. The
matching condition given in expression \ref{eq:matching-condition}
is modified as

\[
\alpha_{G}^{-1}\left(M_{I}\right)-\frac{C\left(G\right)}{12\pi}=\alpha_{G_{i}}^{-1}\left(M_{I}\right)-\frac{C\left(G_{i}\right)}{12\pi}-\frac{\lambda_{i}}{12\pi}\,,\]
where $\lambda_{i}=a_{i}\mathrm{ln}\frac{M_{H}}{M_{I}}$. So the threshold
uncertainty has been included in the matching condition due to presence
of the term involving $\mathrm{ln}\left(M_{H}/M_{I}\right)$.

\begin{table}
\begin{tabular}{ccc}
\hline
\noalign{\vskip3mm}
$\begin{array}{l}
\mathrm{SO(10)\, Higgs}\\
\mathrm{Representation}\end{array}$ & $\begin{array}{l}
\mathrm{Higgs\, multiplets\, contributing\,}\\
\mathrm{to\, threshold\, uncertainty}\\
\left(\mathrm{Decomposed\, under}\, G_{3221D}\right)\end{array}$ & $\left\{ a_{3c},\, a_{2L},\, a_{2R},\, a_{1(B-L)}\right\} $\tabularnewline[5mm]
\hline
\hline
\noalign{\vskip3mm}
$10$ & $\begin{array}{l}
\left(3,\,1,\,1\,-\frac{1}{3}\right)_{10}\oplus\left(\overline{3},\,1,\,1,\,\frac{1}{3}\right)_{10}\\
\left(3,\,1,\,1\,-\frac{1}{3}\right)_{10'}\oplus\left(\overline{3},\,1,\,1,\,\frac{1}{3}\right)_{10'}\end{array}$ & $\left\{ 2,\,0,\,0,\,2\right\} $\tabularnewline[5mm]
\noalign{\vskip3mm}
$16$ & $\begin{array}{l}
\left(3,\,2,\,1,\,\frac{1}{6}\right)_{16}\oplus\left(\overline{3},\,2,\,1,\,-\frac{1}{6}\right)_{\overline{16}}\\
\left(\overline{3},\,1,\,2,\,-\frac{1}{6}\right)_{16}\oplus\left(3,\,1,\,2,\,\frac{1}{6}\right)_{\overline{16}}\end{array}$ & $\left\{ 4,\,3,\,3,\,1\right\} $\tabularnewline[5mm]
\noalign{\vskip3mm}
$45$ & $\begin{array}{l}
\left(3,\,2,\,2,\,-\frac{1}{3}\right)_{45}\oplus\left(\overline{3},\,2,\,2,\,-\frac{1}{3}\right)_{45}\\
\left(8,\,1,\,1,\,0\right)_{45}\end{array}$ & $\left\{ 7,\,6,\,6,\,4\right\} $\tabularnewline[5mm]
\noalign{\vskip3mm}
$54$ & $\begin{array}{l}
\left(6,\,1,\,1,\,-\frac{2}{3}\right)_{54}\oplus\left(\overline{6},\,1,\,1,\,\frac{2}{3}\right)_{54}\\
\left(1,\,3,\,3,\,0\right)_{54}\\
\left(8,\,1,\,1,\,0\right)_{54}\end{array}$ & $\left\{ 8,\,6,\,6,\,8\right\} $\tabularnewline[5mm]
\hline
\end{tabular}

\caption{\label{tab:Threshold-unification}Threshold contribution at the unification
scale}

\end{table}

To avoid any over estimation of the threshold uncertainty we assume
that all the Higgs multiplets, belonging to a single common irreducible
Higgs representation of $SO(10)$, becoming heavy at a given breaking
scale will have the same mass scale around the breaking scale.

The threshold uncertainty at the breaking scale of gauge group $G_{3211}$is
vanishing. The Higgs multiplets, coming from different $SO(10)$ irreducible
Higgs, contributing to the threshold uncertainty at remaining two
intermediate scales, the left-right breaking scale and the unification
scale, are listed in the table \ref{tab:Threshold_left-right} and
\ref{tab:Threshold-unification}, respectively. The corresponding
calculated beta-coefficents, $\left(a_{i}\right)$'s, which include
the contribution from all the Higgs multiplets coming from the same
$SO(10)$ irreducible representation (as their masses are assumed
to be same), are also shown for the two breaking scales.

\begin{figure}
\begin{centering}
\epsfig{file=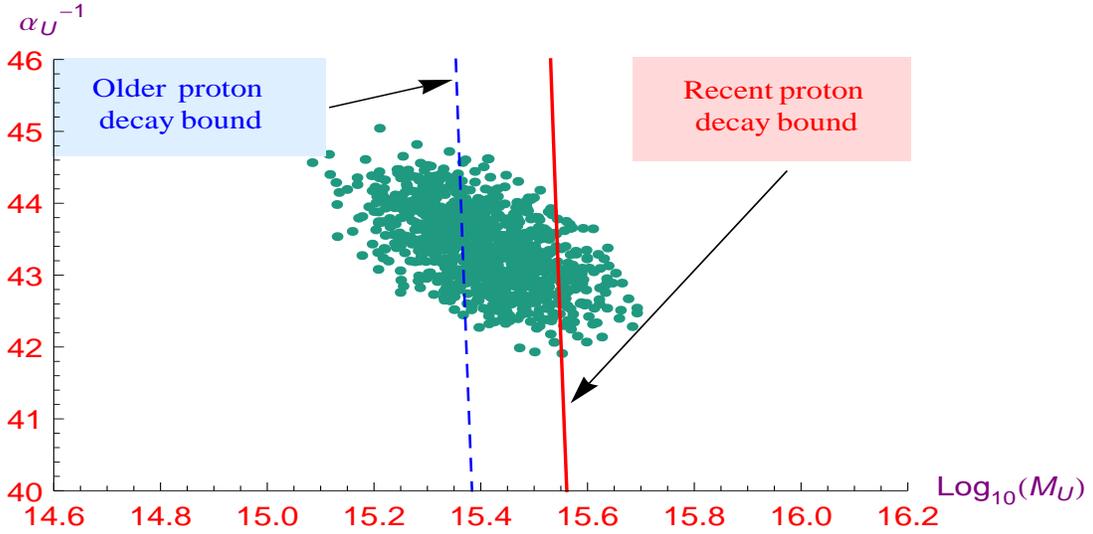, width=0.8\textwidth,height=0.3\textheight}
\par\end{centering}

\caption{\label{fig:Proton-decay-bound}Threshold uncertainty in the unificaton
scale.}

\label{evolution}
\end{figure}

Now using these calculated $a_{i}$'s and including uncertainty in
$M_{H}/M_{I}$, as discussed before, we have shown a scatterd-plot
between coupling constant $\alpha_{U}^{-1}$ and the corresponding
unification scale $M_{U}$ in figure \ref{fig:Proton-decay-bound}.
We have numerically obtained the values for $\alpha_{U}^{-1}$ and
$M_{U}$ for randomally chosen values for $M_{H}/M_{I}$ between the
range $\left(\frac{1}{30}-30\right)$. The random values for all the
Higgs multiplets belonging to the same $SO(10)$ ireducible Higgs
are taken to be same at one perticular breaking scale but different
at the other breaking scale.

Moreover, we have aslo plotted the curve corersponding to the most
recent proton decay bound (red solid curve) \citep{:2009gd} and relatively
older proton decay bound (blue dashed curve) \citep{Shiozawa:1998si}
in figure \ref{fig:Proton-decay-bound} to show the allowed region
in $\alpha_{U}^{-1}$-$M_{U}$ plane. Only the right part of the curve
is allowed by the bound. It is worth noting that the allowed parameter
space is more and more constrained as more updated data on proton
decay bound is available. However, we get a resonable allowed region
in the figure \ref{fig:Proton-decay-bound}, although small, even
after allowing the most conservative threshold uncertainty. So we
expect our model to be satisfactory within the tolerable amount of
threshold uncertainty as far as proton decay bound is concerned.

\section*{Yukawa Sector And Neutrino Masses}

In the present section, we discuss the origin of neutrino masses in
the model. Before proceding further we would like to make it clear
that the discussion about neutrino masses in the present section will
only move around the left-right symmetric model with few inputs from
the $SO(10)$ GUT in motivating about certain patterns for taken Dirac
mass matrices for fermions in our analysis. Moreover, the discussion
will be mainly focused on the matrix structure of low energy neurino
mass matrix allowed with certain assumptions. We will aslo argue,
in what follows, that the consistent neutrino mass spectrum is not
possible within picture of one or two SO(10) singlet fermions $S$.
We start by writing the Yukawa sector of the model as
 \begin{eqnarray}
{\mathcal{L}}_{Y} & = & Y_{ij}~\overline{\ell_{Li}}~\ell_{Rj}\Phi+Y_{ij}'~\overline{\ell_{Li}}~
\ell_{Rj}\Psi+\left(F_{L}\right)_{in}\overline{S_{Rn}}~\ell_{Li}\chi_{L}+\left(F_{R}\right)_{in}\overline{S_{Ln}^{c}}\ell_{Ri}~\chi_{R}\\
 & + & \frac{1}{2}M_{mn}\eta\overline{{S^{c}}_{Lm}}S_{Rn}\end{eqnarray}
 The Yukawa couplings $Y$ and $Y'$ are $3\times3$ matrix, while
$F_{L}$ and $F_{R}$ are $3\times n$ matrices, if we assume that
there are $n$ singlet fermions $S$. So $M$ is a $n\times n$ matrix.
Our study of consistent embedding of the model in $SO(10)$ GUT requires
same structure for both $F_{L}$ and $F_{R}$ up to the scale of left-right
symmetry breaking which, after RG running, can produce small difference
at the weak scale. For the present discussion we assume it to be small
enough so that it can be safely ignored.

The Dirac masses for all the SM fermions including neutrinos are generated
form the the first two terms by giving $vev$ to the bi-doublets as
in any other left-right symmetric model. Since $\Phi$ and $\Psi$
are coming from two independent and real $SO(10)$ 10-dimensional
Higgs, the Dirac mass matrix for neutrinos and charged leptons are
independent. However, the Dirac mass matrix for the up-type quarks
have the same structure as the Dirac mass matrix for the neutrinos
and similarily the Dirac mass matrix for the down-type quarks will
have similar structure as the Dirac mass matrix for the charged leptons
(simply because all SM fermions are assigned to a multiplet of $SO(10)$
GUT). Although, these similarities in the structures are exact only
at the GUT scale, we expect some of its features to be more or less
same even at the low scale. So we can well assume that the Dirac mass
matrix of the neutrinos would almost appear diagonal in the basis
where the charged lepton mass matrix is diagonal. The assumption is
based on the observation that the up-type and down-type quarks are
simultaneously diagonal in the a basis as the quark mixing matrix
is very close to unity. So we borrow the pattern from the quark sector
to the lepton sector where the structure of Dirac mass matrix of the
neutrinos is not directly known unless neutrinos are Dirac fermions.
We expect the following pattern of the Dirac mass matrix of neutrinos
in the diagonal basis of the charged leptons

\[
M_{\nu D}=vY_{lepton}\left(\frac{m_{t}}{m_{b}}\right)=\left(\begin{array}{ccc}
m_{e} & 0 & 0\\
0 & m_{\mu} & 0\\
0 & 0 & m_{\tau}\end{array}\right)\left(\frac{m_{t}}{m_{b}}\right)\simeq v~\left(\begin{array}{ccc}
0.0001 & 0 & 0\\
0 & 0.02 & 0\\
0 & 0 & 0.3\end{array}\right),\]
 where $m_{t}$ and $m_{b}$ are masses of top and bottom quarks and
$m_{e}$, $m_{\mu}$, $m_{\tau}$ are masses of electron, muon and
tau leptons.

The part of the Lagrangian relevant for the neutrino mass generation
is given as follows, \begin{eqnarray}
{\mathcal{L}}_{\nu~mass} & = & \left(\begin{array}{ccc}
\nu , & N^{c} , & S\end{array}\right)_{L}.~X~.\left(\begin{array}{c}
\nu\\
N^{c}\\
S\end{array}\right)_{L}+H.C.\\
 & = & \left(\begin{array}{ccc}
\nu_{i} , & N_{i}^{c} , & S_{m}\end{array}\right)_{L}\left(\begin{array}{ccc}
0 & Y_{ij}v & F_{in}v_{L}\\
\left(Y_{ij}\right)^{T}v & 0 & F_{in}v_{R}\\
F_{mj}^{T}v_{L} & F_{mj}^{T}v_{R} & M_{mn}u\end{array}\right)\left(\begin{array}{c}
\nu_{j}\\
N_{j}^{c}\\
S_{n}\end{array}\right)_{L}+H.C\end{eqnarray}
 Our first task is to analyze the mass spectrum provided by the matrix
$X$ in case of one generation of all fermions. We write the eigenvalue
equation as (eigenvalue: $\lambda$):

\[
\lambda^{3}-Mu~\lambda^{2}-F^{2}v_{R}^{2}\lambda-2YF^{2}vv_{L}v_{R}-MY^{2}uv^{2}=0\]

\textbf{Case 1: $\lambda>>v$,} we get \[
\lambda\left(\lambda+Fv_{R}\right)~\left(\lambda-Fv_{R}\right)=0\]
 The above eigenvalue equation predicts two TeV scale Majorana fermions.
The massless solution contradicts with the condition we started with,
and so is unphysical.

\textbf{Case 2: $\lambda<<v$,} we get 
  \begin{equation}
\lambda=-\frac{2Yvv_{L}}{v_{R}}+\frac{MY^{2}uv^{2}}{F^{2}v_{R}^{2}}\label{seesaw}\end{equation}
 which is of order of eV. So the two Majorana fermions pick up masses
of the order as high as TeV and one remains sufficiently light ($\sim$ eV)
to be identified as light neutrino.

\begin{figure}
\begin{centering}
\epsfig{file=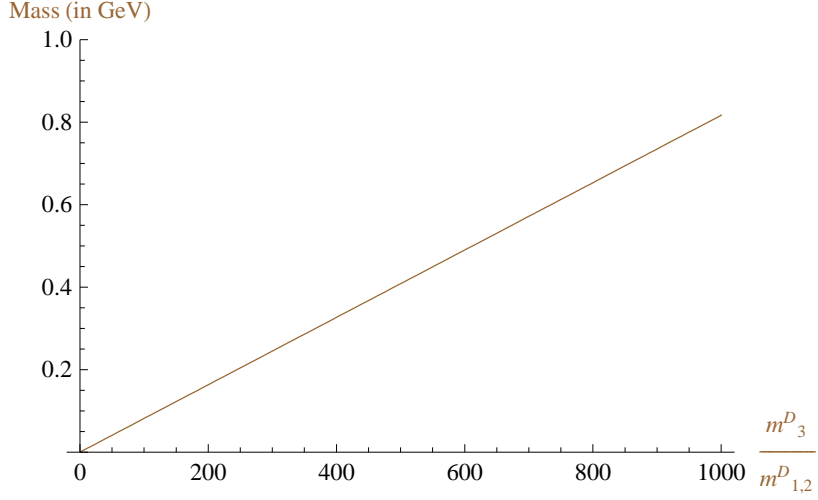, width=0.6\textwidth}
\par\end{centering}

\caption{\label{fig:Variation-of-mass}Variation of mass }

\end{figure}

To make the discussion some more general, we take three generations
for all the SM fermions including the left and right handed neutrinos
but only one generation for the singlet $S$. We look for a possibility
whether it can account for the existing picture of three light active
neutrinos. To search for any such possibility, we try to find out
the mass spectrum, within this scenario, by solving for the eigenvalues
of the matrix $X$. To simplify further, we take all the eigenvalues
of the matrix $M_{\nu D}$ to be same with a common value equal to
the largest one for initial analysis. This enable us to factor out
$\left(\lambda^{2}-z^{2}v^{2}\right)^{2}$ from the algebraic expression
of $\mathrm{Det}\left(X\right)$ predicting four Majorana fermions
of scale around $10$ GeV. The rest of the factors have got the same
form as the expression of determinant in case of one generation of
all SM fermions, as discussed earlier, leading to the two TeV and
one eV scale Majorana fermions. The scenario provides us only one
light neutrino and, hence, can not account for the observed neutrino
mass spectrum. To explore the effect of some possible hierarchy present
in the eigenvalues of the Dirac mass matrix of the neutrino like one
present in the charged lepton mass matrix, we take two of the eigenvalues
to be same and vary their scale below the third one. We are still
able to explicitly get two of the Majorana fermions having mass scale
equal to $m_{e}\left(\frac{m_{t}}{m_{b}}\right)$. One may think that
the remaining two Majorana fermions might get mass scale as light
as eV leading to three light neutrinos. To rule out any such possibility,
we have plotted the masses of the two remaining Majorana fermions
(which comes out to be same) with the ratio of the two mass scales
of the eigenvalues of the Dirac mass mass matrix of the neutrinos
in figure \ref{fig:Variation-of-mass}. We find that the masses do
not go below the lightest mass scale of the eigenvalues of $m_{\nu D}$.
Even in two generation scenario of $S$ fermions, there is not much
progress except we get two eV scale Majorana fermions which is still
not sufficient.

We now turn to the case of three generation for $S$ fermions. One obviously expects to 
get the three light neutrinos. The
basic way to get the low energy neutrino mass matrix has been outlined
in \citep{doub1} which is given as

\begin{eqnarray}
m_{\nu} & = & -\left(\frac{vv_{L}}{v_{R}}\right)\left(Y+Y^{T}\right)+\left(\frac{uv^{2}}{v_{R}^{2}}\right)Y\left(FM^{-1}F^{T}\right)^{-1}Y^{T}~,\nonumber \\
 & = & -\left(\frac{vv_{L}}{v_{R}}\right)\left[\left(Y+Y^{T}\right)+rY\left(FM^{-1}F^{T}\right)^{-1}Y^{T}\right]~,\label{eq:typeIIIformula}\end{eqnarray}
 as we have $uv^{2}=r~vv_{L}v_{R}$ in our model (expression \ref{vlvr1}
) where $r=\left(\lambda_{\chi}-g_{\chi}\right)/h_{\chi}$.

The first term is the type-III seesaw contribution \citep{flhj1988}
and the second term is the double seesaw contribution. With the choice
of the $vevs$, it is obvious that this scenario provides us with
three eV neutrinos.

Now we will try to explore the limits of the expression \ref{eq:typeIIIformula}
for low energy neutrino mass matrix to check its consitency with current
data on neutrino masses and mixing by allowing some very simple form
for matrix $M$. In the basis where charged lepton mass matrix is
diagonal, the neutrino mixing matrix $\left(U_{PMNS}\right)$ is just
the matrix that diagonalizes the $m_{\nu}$:

\[
\left(U_{PMNS}\right)^{T}m_{\nu}U_{PMNS}=m_{\nu}^{Diag}=Diag\left(m_{1},m_{2},m_{3}\right)\,.\]
 The $U_{PMNS}$ mixing matrix is usually parametrized in the literature
as

\[
U_{PMNS}=R_{23}\left(\theta_{23}\right)R_{13}\left(\theta_{13},\delta\right)R_{12}\left(\theta_{12}\right).Dag\left(e^{i\eta_{1}},e^{i\eta_{2}},1\right)\,,\]
 where $R_{ij}$ are the rotation matrices in the $ij$ plane with
angle $\theta_{ij}$. $\delta$ is the CP violating phase associated
with 1-3 rotation and $\eta$'s are the Majorana phases appearing
only in the case of Majorana neutrinos. To date, two mass square differences
and three angle have been measured while CP violation is completely
unknown in the leptonic sector. We take the following observed values
for three mixing angle and two mass square differences at 90$\%$
confidence level from particle data group \citep{pdg} as:

\begin{eqnarray*}
\Delta m_{21}^{2}=m_{2}^{2}-m_{1}^{2} & = & \left(8.0\pm0.3\right)\times10^{-5}\,\mathrm{eV^{2}}\\
\Delta m_{21}^{2}=m_{2}^{2}-m_{1}^{2} & = & 1.9~\mathrm{to}~3.0\times10^{-3}\,\mathrm{eV^{2}}\\
sin^{2}\left(2\theta_{12}\right) & = & 0.86_{-0.04}^{+0.03}\\
sin^{2}\left(\theta_{23}\right) & > & 0.92\\
sin^{2}\left(\theta_{13}\right) & < & 0.19\end{eqnarray*}
 We will be mainly using the mean values of the observed parameters
in our analysis.

In its most general form, it is straight forward to argue that $m_{\nu}$
can accommodate the existing data on neutrino masses and mixing simply
due to the presence of enough number parameters in $F$ and $M$ unless
type III term dominates significantly. An interesting thing would
be to consider some simpler form of the neutrino mass matrix by reducing
appropriate number of parameters with some tolerable assumptions.
The basic idea is to explore the possibility of any such simpler structure
in light of the current neutrino oscillation data.

We start with the assumption that the three singlet fermions $S$
are blind to their generation within themselves leading to the following
democratic structure of matrix $M$ :

\[
M=\left(\begin{array}{ccc}
1 & 1 & 1\\
1 & 1 & 1\\
1 & 1 & 1\end{array}\right)u\]
 The structure allows us to believe that there is no induced mixing
between the left-right neutrinos and the singlets. So, $F$ matrix
can be written as product of a unitary matrix and a diagonal matrix.
The unitary matrix connects the basis of the democratic structure
to the basis where the charged lepton mass matrix becomes diagonal.
To get some more simplicity, we are driven to assume that the two
basis are identical, i.e., the unitary mass matrix is identity matrix.
It leads to the following structure of the low energy neutrino mass
matrix:

\[
m_{\nu}=\frac{vv_{L}}{v_{R}}\left(\begin{array}{ccc}
\alpha^{2}-2\frac{m_{t}}{m_{b}}m_{e} & \alpha\beta & \alpha\gamma\\
\alpha\beta & \beta^{2}-2\frac{m_{t}}{m_{b}}m_{\mu} & \beta\gamma\\
\alpha\gamma & \beta\gamma & \gamma^{2}-2\frac{m_{t}}{m_{b}}m_{\tau}\end{array}\right)\,,\]
 where $\alpha$, $\beta$ and $\gamma$ are the final parameters
appearing in the neutrino mass matrix after absorbing all the parameters
present in $F$, $M$ and $Y$. We take the following familiar tri-bimaximal
form of \citep{Harr} of the $U_{PMNS}$ mixing matrix for our discussion
and attempt to diagonalize $m_{\nu}$ having above structure:

\[
U_{PMNS}=U_{tbm}=\frac{1}{\sqrt{6}}\left(\begin{array}{ccc}
2 & \sqrt{2} & 0\\
-1 & \sqrt{2} & \sqrt{3}\\
1 & -\sqrt{2} & \sqrt{3}\end{array}\right)\,,\]
 where $\theta_{23}=\pi/4$, $\theta_{13}=0$, and $\sin^{2}\theta_{12}=1/3$.

We attempt to diagonalize $m_{\nu}$ with the tri-biamaximal form
of the mixing matrix which requires the following relation of the
parameters $\alpha$, $\beta$ and $\gamma$ with masses of the charged
leptons as:

\begin{eqnarray*}
\alpha & = & 0\\
\beta & \simeq & \frac{m_{\mu}}{\sqrt{\frac{m_{b}}{2m_{t}}\left(m_{\tau}+m_{\mu}\right)}}\simeq0.05\\
\gamma & \simeq & -\frac{m_{\tau}}{\sqrt{\frac{m_{b}}{2m_{t}}\left(m_{\tau}+m_{\mu}\right)}}\simeq-0.75\end{eqnarray*}
 The diagonal neutrino mass matrix comes out of the form:

\[
m_{\nu}^{Daig}\simeq-\frac{2m_{t}}{m_{b}}\left(\begin{array}{ccc}
m_{e} & 0 & 0\\
0 & m_{e} & 0\\
0 & 0 & 2\frac{m_{\mu}m_{\tau}}{\left(m_{\mu}+m_{\tau}\right)}\end{array}\right)\left(\frac{vv_{L}}{v_{R}}\right)\]
 So the present form of $m_{\nu}$ and $U_{PMNS}$ produces degenerate
masses for the two light neutrinos which is likely to be cured once
we slightly deviate from tri-bimaximal form of $U_{PMNS}$. The deviation
can be realized either by taking non-maximal value of $\theta_{23}$
or non vanishing value of $\theta_{13}$ or both. We take only non-zero
value of $\theta_{13}$ to be the sole realization of the deviation
for our purpose. The deviated form of tri-baimaximal matrix for very
small value of $\theta_{13}$ can be parametrized as:

\[
U_{PMNS}=\frac{1}{\sqrt{6}}\left(\begin{array}{ccc}
2 & \sqrt{2} & \theta_{13}\\
-1-\sqrt{2}\theta_{13} & \sqrt{2}-\theta_{13} & \sqrt{3}\\
1-\sqrt{2}\theta_{13} & -\sqrt{2}-\theta_{13} & \sqrt{3}\end{array}\right)\]

While trying to diagonalize the $m_{\nu}$, numerical methods are
used to find out the desired values of the free parameters. We find
that the degeneracy encountered in the case of tri-bimaximal mixing
matrix disappears as soon as finite value of $\theta_{13}$ is introduced.
This finite value is determined by imposing the condition $\Delta m_{21}^{2}/\Delta m_{31}^{2}\simeq0.033$
which leads to following value of $\sin\theta_{13}$

\[
\sin\theta_{13}=0.11.\]
 The value is well within the allowed value for $\theta_{13}$ from
oscillation data. The correct scale of the mass square differences
is easily achieved by adjusting the over all scale of the neutrino
mass matrix. The corresponding values of the other parameters come
out to be \begin{eqnarray*}
\alpha & = & 0.02\\
\beta & = & 0.06\\
\gamma & = & -0.75\end{eqnarray*}

The point we would like to emphasize is that even the simple structure
of the mass matrix taken in our analysis is able to account for the
existing framework of three active light neutrinos even though the
assumptions may not correspond to any real underlying symmetry.

\section*{Dark Energy}

We shall now show how the model can accommodate the proposal
of the mass varying neutrinos (MaVaNs) \cite{mavans1,mavans2}.
The basic idea behind the mass varying neutrinos is that some
scalar field, the acceleron, acquires a value of the order of
$10^{-3}$~eV, which gives an effective potential that contributes
to the dark energy with the equation of state $\omega = -1$.
However, till recently the neutrino masses were contributing
to the effective potential much more strongly and the combined
fluid of the background neutrinos and the accelerons were
behaving as dark matter with the equation of state $\omega = 0$.
As the neutrino masses were varying with time, the contribution
of the background neutrino density to the effective potential
were changing. Only in the recent past, the contribution of
the acceleron field to the effective potential became stronger
than the background neutrinos, changing the equation of state
of the combined fluid, and the universe started
accelerating with dark energy domination. This can then
explain why the scale associated with the amount of dark energy
is comparable to the neutrino masses, why the amount of
dark energy is comparable to the ordinary matter, and why the
universe is dominated by dark energy only now and for the
rest of the time in the past the evolution of the universe was
governed by matter.

In spite of these advantages, the MaVaNs
scenario are not free of problems. We shall now try to
explain how the MaVaNs scenario
can be accommodated in a grand unified theory. After
describing the generic features of the MaVaNs, following
the original proposal \cite{mavans2}, we shall explain
how our present model answer this question. We shall not
restrict ourselves to any particular choice for the
acceleron field, and hence, consider
the potential for the acceleron field to be same as that
considered in the original proposal. At the end
we shall mention how the present model can be extended to
allow a milli-eV mass pseudo-Nambu-Goldstone Boson (pNGB), which
can become the acceleron field.

We shall now mention this possible origin of the acceleron field
in an extension of our model. Following the prescription
followed in ref. \cite{pas}, we introduce three $\eta$ and several
Higgs doublets. The $vevs$ of the fields $\eta$ would then give
rise to global symmetries, which are allowed by all the Yukawa
couplings due to the choice of quantum numbers of the Higgs doublets
under these global symmetries.
However, when the Higgs doublets acquire $vevs$, the global symmetries
will be broken and there will be pNGBs, which couple to the
neutrino masses. Although the dynamics of the pNGBs are not
specified, the masses and the potentials of the pNGBs are
determined by the Coleman-Weinberg potential, as demonstrated in ref. \cite{pas}.
Since the introduction of several Higgs doublets may not allow the
the gauge coupling unification, we shall not discuss this extension
any further. Moreover, there could be some other origin of the
acceleron field, so from now on we shall only mention the generic features
of this model.

In a generic MaVaNs models, the coupling between neutrino mass and
$\aa$ induces the following effective potential
\begin{equation}
V=\left(\rho_{\nu}-3P_{\nu}\right)+V_{0}(m_{\nu})
\end{equation}

\noindent Here the scalar potential $V_{0}(m_{\nu})$ is due to the
acceleron field (written as a function of neutrino mass) and $P_{\nu}$
is pressure of the neutrino fluid. In the late time evolution the
non-relativistic limit i.e. $m_{\nu}\gg T$ is of particular interest.
In this case $P_{\nu}\sim0$ and one can write the effective potential
as, \begin{equation}
V=m_{\nu}n_{\nu}+V_{0}(m_{\nu})\,.\end{equation}

\noindent The acceleron field will be trapped at the minima of the
potential, which ensures that as the neutrino mass varies, the value
of the acceleron field will track the varying neutrino mass. One can
write equation of state in the non-relativistic case for a combined
fluid of neutrino $+$ acceleron; \begin{equation}
w\,=\, P/\rho\,=\,\frac{P_{\aa}}{m_{\nu}n_{\nu}+\rho_{\aa}}\end{equation}

\noindent One generic feature of this solution is that it gives $\omega\approx-1$
at present. The most important feature of this scenario is that the
energy scale for the dark energy gets related to the neutrino mass,
which is highly desirable. This also explains why the universe enters
an accelerating phase now \citep{wet1}.

We shall now discuss the implementation of the $\nu$DE~mechanism in
our model. For simplicity, we consider only one-generation scenario.
The effective scalar field potential of the scalar is of the Coleman-Weinberg
type i.e.
\begin{equation}
 V_{0}  = \Lambda^{4}~\log(1+|{M_{s}(\aa)/\bar{\mu}|}
\end{equation}
\noindent
where, $M_{s}$ is the singlet fermion mass.
We assume that $M_{s}(\aa)/\bar{\mu}\gg 1$.  $$M_{s}=M\langle\eta\rangle=M~u$$ depends
on the acceleron field $\aa$. Thus the neutrino mass becomes
a dynamical quantity. When the neutrinos become non-relativistic the
dependence of $M_{s}$ on $\aa$ governs the dynamics of the dark
energy.
\noindent
$\Lambda$  is chosen in such a way to yield the dark energy
density $\Omega_{\text{DE}}\approx 0.7$.
This type of potentials are extensively used in the dark energy
literature \cite{mavans1,stab}. Now we can write the effective low-energy
Lagrangian in our model
\begin{equation}
-{\mathcal{L}}_{eff}=M_{s}(\aa){\frac{Y^{2}}{F^{2}}}{\frac{v^{2}}{v_{R}^{2}}}~\nu_{i}\nu_{j}
+H.c.+\Lambda^{4}\log(1+|M_{s}(\aa)/\bar{\mu}|)\,,
\end{equation}
 From the choices we have made about the $vev$s, we have retained
only the dominant double seesaw term \ref{seesaw} in the effective
Lagrangian. As $u\sim O(eV)$, the mass parameter $M_{s}$ is of the
order of eV. Since the ratio $(v/v_{R})^{2}\sim10^{-2}-10^{-3}$,
the Yukawa couplings coupling to be of order unity. Thus the first
two terms in equation (14) are comparable to the last term describing
the dark energy potential.

The Majorana mass of neutrino varies with the acceleron field through
the parameter $M_{s}$ and the mass scale of this parameter remains
near the scale of dark energy \emph{naturally}. The interesting feature
of our model is that we do not need any unnaturally small Yukawa couplings
or symmetry breaking scale to achieve this naturalness requirement.
Also the variation of $M_{s}$ does not affect charged fermion masses
in the model. Moreover, the electroweak symmetry breaking scale $v$
and the $U(1)_{R}$ breaking scales are comparable and hence the new
gauge boson corresponding to the group $U(1)_{R}$ will have usual
mixing with $Z$ and should be accessible at LHC.

Since the local minimum of the potential relates the neutrino mass
to a derivative of the acceleron potential, the value of the acceleron
field gets related to the neutrino mass. The acceleron field provide
an effective attractive force between the neutrinos. When this
effective force is stronger than the gravity, perturbations in the
neutrino-acceleron fluid become unstable. The source of the free-energy
comes from the attractive interaction between the neutrino and the
acceleron field. The instability is similar to that of the Jeans instability
found in a self-gravitating system. The instability can lead to inhomogeneity
and structure formation; the instability would grow till the degeneracy
pressure of the neutrinos would arrest the growth. The final state
of the instability would produce neutrino lumps or nuggets \cite{stab,wet2}.
The neutrino lumps would then behave as dark matter and will not affect
the dynamics of the acceleron field \cite{wet2}. This instability
is a generic feature of MaVaNs scenario, however it can be suppressed
if the neutrino become superfluid \cite{bhsar} or if the MaVaNs
perturbations become non-adiabatic.


\section*{Conclusions}

We have constructed a left-right symmetric model of $\nu DE$ that
can be embedded in an $SO(10)$ GUT. After discussing the Higgs content
needed for the model, details of potential minimization have been
carried out considering all possible allowed terms. In particular,
we have tried to explore the possibility of choosing the minima such
that only neutral Higgs components get $vev$ without constraining
the couplings constants. But it turns out that some such constraints
are needed in most general form of the potential. The complete analysis
allows the desired ordering of the $vev$s. Then we study the embedding
of this left-right symmetric model in $SO\left(10\right)$ GUT. We
show that $SO\left(10\right)$ GUT with Higgs multiplets $S(54)$,
$A(45)$, two $H(10)$, $C(16)\oplus\overline{C(16)}$, $\eta(1)$
along with an additional fermion singlet is able to accommodate the
left-right symmetric model. The embedding allows the Pati-Salam and
the left-right symmetry group breaking scales to be different by orders
of magnitudes. We have studied the one loop RG running of various
couplings constant and have found that the desired assignment for
$vev$ values for different Higgs fields is consistent with the gauge
unification. Then the origin and possible structure of neutrino masses
and matrix have been discussed in detail. It has been shown that generation
of three light active neutrinos of $eV$ scale is not possible in
scenario with one or two $SO(10)$ singlets fermions. In the generic
case of three singlets, we have taken a simple structure of neutrino
mass matrix with some tolerable assumptions and shown that the structure
is consistent with current data on neutrino masses and mixing. Then
we described implementation of $\nu DE$ in the model. The model allows
the mass parameter of the singlet, which varies with the acceleron
field, to have the same scale as the scale of dark energy satisfying
the desired naturalness requirement.

\end{document}